\documentclass[iop]{emulateapj}
\shorttitle{SN 2006bc}
\shortauthors{Gallagher et al.}
\usepackage{booktabs}
\usepackage{natbib}

\begin{document}

\title{Optical and Infrared analysis of Type II SN 2006bc}

\author{Joseph S. Gallagher\altaffilmark{1}, B.E.K. Sugerman\altaffilmark{3}, Geoffrey C. Clayton\altaffilmark{2},  J.E. Andrews\altaffilmark{2}, J. Clem\altaffilmark{2}, M.J. Barlow\altaffilmark{4}, 
B. Ercolano\altaffilmark{5}, J. Fabbri\altaffilmark{4}, M. Otsuka\altaffilmark{6-7}, R. Wesson\altaffilmark{4}, and M. Meixner\altaffilmark{7}}

\altaffiltext{1}{MPCS Department, University of Cincinati, Blue Ash College, 9555 Plainfield Rd., Blue Ash, OH 45236; gallagjl@uc.edu}
\altaffiltext{2}{Department of Physics and Astronomy, Louisiana State University, 202 Nicholson Hall, Baton Rouge, LA 70803; gclayton@fenway.phys.lsu.edu, jandrews@phys.lsu.edu, jclem@phys.lsu.edu}
\altaffiltext{3}{Department of Physics and Astronomy, Goucher College, 1021 Dulaney Valley Rd., Baltimore, MD 21204; ben.sugerman@goucher.edu}
\altaffiltext{4}{Department of Physics and Astronomy, University College London, Gower Street, London WC1E 6BT, UK; mjb@star.ucl.ac.uk, jfabbri@star.ucl.ac.uk, rwesson@star.ucl.ac.uk}
\altaffiltext{5}{Universit$\ddot{a}$ts-Sternwarte M$\ddot{u}$nchen, Scheinerstr. 1, 81679 MŸnchen, Germany; ercolano@usm.uni-muenchen.de}
\altaffiltext{6}{Institute of Astronomy and Astrophysics, Academia Sinica P.O. Box 23-141, Taipei 10617, Taiwan, R.O.C.;
otsuka@asiaa.sinica.edu.tw} 
\altaffiltext{7}{Space Telescope Science Institute, 3700 San Martin Drive, Baltimore, MD 21218; otsuka@stsci.edu, meixner@stsci.edu}      

\begin{abstract}
We present nebular phase optical imaging and spectroscopy and near/mid-IR imaging of the Type II SN 2006bc.  Observations reveal the central wavelength of the symmetric H$\alpha$ line profile to be red-shifted with respect to the host galaxy H$\alpha$ emission by day 325.  Such an phenomenon has been argued to result from an asymmetric explosion in the iron-peak elements resulting in a larger mass of $^{56}$Ni and higher excitation of hydrogen on the far side of the SN explosion.  We also observe a gradual blue-shifting of this H$\alpha$ peak which is indicative of dust formation in the ejecta.  Although showing a normal peak brightness, V $\sim$ -17.2, for a core-collapse SN, 2006bc fades by $\sim$6 mag during the first 400 days suggesting either a relatively low $^{56}$Ni yield, an increase in extinction due to new dust, or both.  A short duration flattening of the light curve is observed from day 416 to day 541 suggesting an optical light echo.  Based on the narrow time window of this echo, we discuss implications on the location and geometry of the reflecting ISM.  With our radiative transfer models, we find an upper limit of 2 x 10$^{-3}$ M$_{\odot}$ of dust around SN 2006bc.  In the event that all of this dust were formed during the SN explosion, this quantity of dust is still several orders of magnitude lower than that needed to explain the large quantities of dust observed in the early universe. 
\end{abstract}


\section{Introduction}
The majority of dust formed in galaxies at the current epoch originates in the stellar winds of low mass, asymptotic giant branch (AGB) stars \citep{1989IAUS..135..445G}.  Although the effective temperature of AGB stars, around 3000K, exceeds the condensation temperature for most compositions of dust grains, pulsation or convection driven shock waves propagating radially from the star can create transitory dense, cool layers that are suitable for dust formation.  Once created the newly formed dust can be ushered away into the interstellar medium (ISM) via radiation pressure.

Recent sub-mm observations of high redshift quasi-stellar objects (QSO's) and galaxies have revealed large quantities of dust at z $>$ 5 \citep{2003A&A...409L..47B, 2007ApJ...662..927D}.  Such redshifts correspond to an age of the Universe of 1 Gyr or less, considerably younger than the expected delay time for dust formation in AGB stars \citep{2010A&A...522A..15M}.  An alternate mechanism for the creation of this high redshift dust is dust formation in the ejecta of supernovae (SNe) given that the delay time associated with SNe is measured in millions, not billions, of years.  This was first suggested as a possible mechanism by \citet{1967AnAp...30.1039C}.  Recent theoretical models predict that 0.1-1 M$_{\sun}$ of dust should form in the ejecta of every SN event \citep{2001MNRAS.325..726T,2003ApJ...598..785N}; a production rate that can certainly account for the large amount of dust at high redshift \citep{2003MNRAS.343..427M,2006MmSAI..77..643M,2007ApJ...662..927D}.  Observational evidence of SN dust formation was first seen within the ejecta of SN 1987A \citep{1991supe.conf...82L,1993MNRAS.261..535M, 2007MNRAS.375..753E}.  This evidence came in the form of (1) the development of an infrared excess, (2) an increase in the optical extinction, and (3) the appearance of blue-shifted emission line profiles due to the greater attenuation of emission from the far-side, red-shifted ejecta emission. 


To date, evidence of dust formation has been seen in all types of core-collapse SNe, save for type Ic, and no signs of dust formation have been observed in thermonuclear, SNe Ia.  In addition to SN 1987A, dust has been shown to have condensed in the ejecta of five Type II SNe, 1990I \citep{2004A&A...426..963E}, 1999em \citep{2003MNRAS.338..939E}, 2003gd \citep{2006Sci...313..196S,2005MNRAS.359..906H,2007ApJ...665..608M}, 2004et \citep{2006MNRAS.372.1315S,2009ApJ...704..306K, jo_2004et_paper, ben_2004et_paper}, and 2005af \citep{2006ApJ...651L.117K}.  In each case the quantity of dust formed was less than 10$^{-3}$ M$_{\sun}$, considerably less than that needed to explain the dust observed at high-z.  More recently, observations of SNe exhibiting strong circumstellar interaction have revealed a secondary location, outside of the ejecta, of dust formation.  These observations have shown evidence of dust formation within the cool, dense shell (CDS, \citet{2003ApJ...594..312D}) between the forward and reverse shocks formed by the interaction of the ejecta and circumstellar material (CSM).  This mechanism has been reported in four SNe, 1998S \citep{2004MNRAS.352..457P}, 2005ip \citep{2009ApJ...695.1334S,2009ApJ...691..650F}, 2006jc \citep{2008ApJ...680..568S,2008MNRAS.389..141M}, 2007it \citep{2011ApJ...731...47A}, and 2007od \citep{2010ApJ...715..541A}.  However, despite their potential to produce dust in both their ejecta and in their CDS, CSM-interacting SNe have been shown only to form comparable amounts of dust to their CSM deficient counterparts.  

Recent studies that have reevaluated the delay time limits for dust formation by AGB stars have concluded that AGB stars may be more efficient dust producers at high redshift than first expected, but they have not been able to rule out SNe as major contributers \citep{2009MNRAS.397.1661V,2011ApJ...727...63D}.  Other possible, non-stellar, sources of dust have been put forward such as the growth of dust grains in the ISM.  \citet{2009ASPC..414..453D}  suggested that although SNe contribute a portion of the dust found in the ISM, that most of the dust mass observed in the ISM comes from the growth of these seed grains after they arrive in the ISM.

In spite of the great interest in understanding the origins of dust at high redshift, there is still no firm conclusion yet as to the ability of SNe to significantly contribute to the overall dust budget in the galaxy.  The number of SNe that have been observed at multiple epochs and in multiple wavelengths is still very small, and the more SNe that are observed, the greater the realization that each SN is unique.  This makes it of paramount importance to increase the sample of observed core-collapse SNe, so that a statistically significant conclusion can be made on their ability to form dust.  To this end, we report multi-temporal, multi-wavelength observations of core-collapse SN 2006bc.  In \S 2 we will describe our observing program and our reduction methods.  In \S3 we will report our results, and we summarize our conclusions in \S4.

\section{Observations \& Reductions}

\subsection{Previous Observations and Relevant Statistics}

SN 2006bc was discovered on 2006 March 24.65 UT approximately 49$\arcsec$ west and 43$\arcsec$ north of the center of its host, NGC 2397 (Figure \ref{fig1}, \citet{2006CBET..446....1M}).  An optical spectrum taken by \citet{2006CBET..450....1P} with the ESO Very Large Telescope on 2006 March 28.06 UT showed narrow Balmer emission superimposed on a blue continuum.  The width of these Balmer lines suggested an expansion velocity of $\sim$2000 km s$^{-1}$.  Unresolved Na I D lines originating within the Milky Way and the host galaxy were also present, and both lines had measured equivalent widths of $\sim$ 1.0 \AA\ \citep{2006CBET..450....1P}.  \citet{2006CBET..450....1P} also determined a B$-$V color for SN 2006bc  of $\sim$ 0.2 while coincident Swift Ultraviolet/Optical (UVOT) observations found U$-$B $\sim$ -0.9 $\pm$ 0.1 \citep{2006ATel..776....1I}.  Based on the color-SN type diagnostics of \citet{2002PASP..114..833P}, SN 2006bc is considerably bluer than most type II SNe with colors closer to that of type Ia SNe.  However, the presence of hydrogen emission and the similarities of its colors to type II-P SN 2005cs and 2006at led to its classification as a type II-P \citep{2006ATel..776....1I}.  

SN 2006bc was further imaged on 2006 October 14 (203 d) with the ACS Wide Field Camera (WFC) in the $F435W$, $F555W$, and $F814W$ filters by \citet{2009MNRAS.395.1409S}.  Alignment of these images with pre-explosion images taken with the Wide Field Planetary Camera 2 (WFPC2) onboard the \textit{Hubble Space Telescope (HST)} on 2001 November 17 resulted in no detection of the progenitor star..  However, based on the detection limit of the pre-explosion images, and assuming the progenitor to be a red supergiant, \citet{2009MNRAS.395.1409S} derived an upper mass limit for the progenitor of SN 2006bc of 12 M$_{\sun}$.  
 
Based on the equivalent width measurements of the unresolved Na I D lines for the NGC 2396 and the Milky Way by \citep{2006CBET..450....1P}, and using with the empirical relation of \citet{1990A&A...237...79B} relating EW(Na I D) with the intrinsic color excess, E(B-V), and R(V) = 3.1, we find a substantial total foreground extinction to SN 2006bc of A$_{\textit{v}}$ $\sim$ 1.6.  According to \citep{2009AJ....138..323T}, NGC 2397 is at a distance of 20.3 Mpc, and we calculated the recessional velocity of NGC 2397 via redshifted HII emission lines in our own optical spectra to be (1462$\pm$12) km s$^{-1}$.  Finally, based on negative detections by \citet{2006CBET..446....1M} as late as 2006 March 16.59 UT, we adopt an explosion date that coincides with the date of discovery (JD 2453819.15).

\subsection{Optical Spectroscopy}

For this program data were taken with the Gemini South telescope utilizing the GMOS-S instrument during the 2007A and 2007B semesters (GS-2007A-Q-4 \& GS-2007B-Q-39).  Dates of observation include 2007 February 12 (324 d), March 18 (358 d), May 11 (412 d), September 20/24 ($\sim$544 d), November 11/14 ($\sim$596 d), and 2008 January 16 (662 d; see Table 1).  During the first three epochs, 3$\times$900s spectra were obtained in long slit mode.  The exposure time was augmented to 3$\times$1050s for the last three epochs to compensate for the expected SN optical decline.  The GMOS-S spectrograph, equipped with the B600-G5323 grating, allowed for 2760 \AA\ coverage and a FWHM resolution of $\sim$3.9 \AA.  The central wavelengths of individual spectra were 5950\AA, 5970\AA, and 5990\AA, respectively, to allow for effective removal of chip gaps and to avoid having the gaps fall on important emission features.  A 0.75$\arcsec$ slit width was used with the B600-G5323 grating in first order along with 2$\times$2 binning in the low gain setting. 

Spectra were reduced using the IRAF\footnote{IRAF is distributed by the National Optical Astronomy Observatory, which is operated by the Association of Universities for Research in Astronomy (AURA) under cooperative agreement with the National Science Foundation.} \textit{gemini} package.  An example of the slit orientation from day 324 is shown in Figure \ref{fig2}.  The inset displays a section of the 2-D spectrum including emission from H$\alpha$ and [NII].  The figure highlights the  contamination of galactic emission in the light from the SN.  The SN and sky regions used during the 1-D spectral extraction were determined by visual inspection.  Care was taken when defining the regions of sky to include continuum emission from the galaxy, but to avoid stars and clumpy HII regions falling on the slit as shown in Figure \ref{fig2}.  The 3 spectra obtained each individual night were averaged to improve signal-to-noise.  

\subsection{Optical Imaging}

Coincident optical imaging was obtained using the GMOS-S imager in three SDSS filters, \textit{g$^{ \prime}$}, \textit{r$^{ \prime}$}, and \textit{i$^{ \prime}$}.  The images were reduced and stacked using the standard routines within the \textit{gemini} package.  To remove as much contamination from galactic light as possible, we performed PSF-matched difference imaging in each filter using the IRAF \textit{difimphot} package \citep{1996AJ....112.2872T,2005ApJS..159...60S}.  The procedure entailed registering and matching the PSFs of an object and reference image.  Given the lack of pre-explosion GMOS-S images of NGC 2397, we opted to use our day 662 \textit{g$^{\prime}$r$^{\prime}$i$^{\prime}$} images as the reference epoch.  In each filter, the reference image was subtracted from the images of the earlier epochs, resulting in a final PSF-matched difference image in \textit{g$^{\prime}$r$^{\prime}$i$^{\prime}$} for days 324, 358, 412, 544, and 596.  An example is shown in Figure \ref{fig3} of our day 324 \textit{i$^{\prime}$} subtraction.  

PSF-photometry was then performed on the difference images using the IRAF \textit{daophot} package.  The instrumental \textit{g$^{\prime}$r$^{\prime}$i$^{\prime}$} magnitudes were converted into the standard Johnson-Cousins $VRI_{c}$ using a transformation involving a least-squares fit with a floating zero-point \citep{2007ApJ...669..525W}. A $BVRI_{c}$ photometric sequence of secondary standards was derived for the 2006bc field.  Observations of the secondary standard stars and $\sim$30 stars from the lists of \citet{2009AJ....137.4186L} were taken with the Y4KCam CCD on the Yale 1.0-m telescope operating at the Cerro Tololo Inter-American Observatory on 2008 October 28.  For full details on the data reduction and the creation of our photometric sequence see Appendix A of \citet{2010ApJ...715..541A}.  The standard stars are shown in Figure \ref{fig1} and their corresponding magnitudes are reported in Table 4 .  PSF-photometry on the non-differenced images of these secondary standard stars allowed us to derive the unique, filter-dependent zero-points for each night.  It is important to note that any remaining SN flux present in our reference epoch will lead to our underestimating the flux at a given epoch.  

Late time imaging data were obtained by \textit{HST} using WFPC2 on 2007 September 09 (532 d) and 2008 February 17 (694 d).  Data were taken using the $F606W$and $F814W$ filters, and the SN was centered on the Planetary Camera (PC) CCD chip.  The images were pipeline reduced, cosmic rays were removed, and images were drizzled to a higher spatial resolution, 0.03$\arcsec$ pixel$^{-1}$ (65\% of the native pixel scale), using the \textit{STSDAS} package.  Instrumental magnitudes were determined via PSF-photometry using the \textit{daophot} package.  We generated simulated PSFs for this analysis using the PSF creation software \textit{Tiny Tim} \citep{tiny_tim}.  The $F606W$ and $F814W$ instrumental magnitudes were transformed into Johnson V and I$_{c}$, respectively, and charge transfer efficiency (CTE) corrections were applied using the iterative procedure detailed in \citet{2000PASP..112.1397D,2009PASP..121..655D}

\subsection{Near-Infrared Imaging}


Wide band $F110W$, $F160W$, and $F205W$ images of 2006bc were obtained on 2007 September 17 (541 d) and 2008 February 17 (694 d) using \textit{HST} equipped with the Near Infrared Camera and Multi-Object Spectrograph (NICMOS, Meixner: GO11229).  The NIC2 camera employing the square wave dither pattern was utilized.  The images were pipeline reduced and photometry was carried out in similar fashion to that of the WFPC2 data.  Model PSFs were created with \textit{Tiny Tim} and were used along with \textit{daophot} to calculate instrumental magnitudes.  \textit{HST} system magnitudes (ST) were calculated from the measured count rate (CR) and the header keywords PHOTFLAM according to Equation (1). 

\small
\begin{equation}
STMAG = -2.5 \times \log_{10}(PHOTFLAM \times CR) - 21.1\\
\end{equation}
\normalsize

PHOTFLAM is defined as the mean flux density in units of erg cm$^{-2}$ s$^{-1}$ \AA$^{-1}$ that produces 1 count per second in a given HST observing mode.  The ST magnitudes were then transformed into $JHK$ magnitudes using the IRAF \textit{synphot} package. 
\\

\subsection{Mid-Infrared Imaging}

Three epochs of \textit{Spitzer}/IRAC (3.6, 4.5, 5.8, and 8.0 $\mu$m) and MIPS (24$\mu$m) images were taken along with two epochs of IRS Peak-Up image (PUI) data at 16$\mu$m.  The IRAC data were taken on 2007 September 12 (537 d), 2008 February 12 (690 d), and 2008 May 08 (776 d).  The IRS PUI data were taken on 2007 August 02 (496 d) and 2008 January 01 (648 d) while the MIPS data were taken on 2007 September 15 (540 d), 2008 February 10 (688 d), and 2008 April 15 (753 d).  The chronology of all observations is summarized in Table 1.  To improve photometric quality, all the data were mosaicked and resampled using the software package MOPEX \citep{2005ASPC..347...81M}.  The IRAC images were drizzled to 0.75$\arcsec$ pixel$^{-1}$ (63\% of the native pixel scale), the IRS PUI 16$\mu$m to 0.9$\arcsec$ pixel$^{-1}$ (50\%), and the MIPS 24$\mu$m to 1.2$\arcsec$ pixel$^{-1}$ (50\%).  Difference imaging analysis was performed as described above with the day 776 (IRAC), day 753 (MIPS), and day 688 (IRS/PUI) images subtracted from the earlier epochs for the three instruments, respectively. PSF-photometry was performed at the SN position using DAOPHOT while a customized implementation of the IRAF package DIMPHOT, called DIPHOT \citep{1996AJ....112.2872T}, was used to obtain photometric errors or upper detection limits for the differenced images.  
 
\section{Results}

\subsection{Optical Emission Line Profiles}
 
As noted earlier, the spectroscopic analysis of SN 2006bc is difficult, particularly at late times, due to its location within the bright SBb host galaxy, NGC 2397.  Our six spectra, stretching from day 325 to 663, are shown in Figure \ref{fig4}.  The spectra exhibit prominent narrow emission lines of H$\beta$, [OIII], [NII], [SII], and H$\alpha$ typically observed in the HII regions of star-forming galaxies.  The spectra also show lines of [OI] and H$\alpha$ with widths consistent with emission from the evolving SN ejecta (500-1000 km s$^{-1}$).  The evolution of [OI] and H$\alpha$ are shown in Figures \ref{fig5} and \ref{fig6}, respectively.  Although Figure \ref{fig5} has been corrected for the velocity of the host galaxy, Figure \ref{fig6} has not.  Little change is observed in the [OI] emission profiles; however, in spite of strong contamination from nebular emission, noticeable evolution is observed in the profile of H$\alpha$.  Figure \ref{fig6} shows the evolution of the H$\alpha$ + [NII] emission profile for which we have carefully aligned the narrow host galaxy H$\alpha$ and [NII] lines.  Apparent is the underlying emission from the SN fading over the course of our observing period.  

In order to separate the nebular emission from the SN emission, we simultaneously fitted Gaussian curves to the blended H$\alpha$ + [NII] profile of the host galaxy and SN.  The results of this analysis are given in Figure \ref{fig7}.  The top two panels show the fit to our day 325 profile on the left and its Gaussian components on the right.  Subsequent plots show the fits to our remaining five epochs.  The most obvious characteristic of the SN emission is that by day 325 the central wavelength of the SN emission is red-shifted with respect to the galactic H$\alpha$ emission by $\sim$800-900 km s$^{-1}$.  A similar phenomenon was observed in the optical spectra of SN1987A, 1999em, and 2004dj \citep{1991supe.conf...36P,2003MNRAS.338..939E,2005AstL...31..792C}.  In the case of SN 1987A, the emission peaks of several elements, including hydrogen, were observed to be redshifted from 600-1000 km s$^{-1}$ during the period 150 to 600 days after explosion.  It was subsequently argued by \citet{1991SvA....35..171C} that this redshift was due to an asymmetry in the $^{56}$Ni ejecta.  Just like SN 2006bc, SN 1987A did not show this asymmetry in the [OI] emission lines, but it did show strong asymmetry in Fe II and Ni I \citep{1989MNRAS.238..193M}.  This led \citet{1991SvA....35..171C} to suggest that the asymmetries involved only the iron-peak elements with the redshifting of the H$\alpha$ line being a secondary effect due to an increase in the excitation of hydrogen in positions where the mass of $^{56}$Ni is higher.  Given that we are observing the emission line profile asymmetries in H$\alpha$, but not [OI], it is likely that we are seeing a similar effect in the optical spectra of SN 2006bc. 

Figure \ref{fig8} shows the evolution of our Gaussian model to the SN H$\alpha$ emission profile with time.  The inset shows the central wavelength evolution of these models which reveals an apparent blueshift of the H$\alpha$ peak from day 325 to day 663.  Figure \ref{fig8} shows a blueshift of approximately 200 km s$^{-1}$ over the course of our observing program.  Although it is a possibility the such a blueshift could be a further consequence of the apparent asymmetrical explosion, blueshifted emission lines are also often the result of the formation of new dust in the ejecta resulting in the emission from the back-side of the explosion experiencing greater attenuation than emission from the front-side.  Consequently, the apparent blueshift of the peak in Figure \ref{fig8} could be a sign of dust formation in the ejecta of SN 2006bc. 

\subsection{Radiative Transfer Modeling}
From our PSF-matched photometry we were successful in obtaining detections in all four IRAC channels on day 537, but we did not detect the SN on day 690 in any channel.  Furthermore, we did not obtain detections in the MIPS or IRS/PUI data at any epoch.  We combined our IRAC detections on day 537, our $\textit{HST}$ results from day 532, and upper limits at 16 $\mu$m (day 496) and 24 $\mu$m (day 688) into a single SED for 2006c approximately 550 days past maximum light.  The SED of SN 2006bc is shown in Figure \ref{fig9}.  The near and mid-IR flux measurements are shown in Table 2) 

Even with mere upper limits beyond 10 $\mu$m, the IRAC results show evidence for strong IR excesses attributable to warm dust.  The dotted-line in Figure \ref{fig9} represents our blackbody fit to the SED.  The data are well fitted by the sum of a 4900 K blackbody, representing the emission from the hot, optically thick, ejecta, and a 500 K modified blackbody due to the emission from a spherical distribution of warm dust with an outer radius of $\sim$570 AU.  The modified blackbody is a normal blackbody subject to a $\lambda$$^{-1}$ emissivity law. 

Quantitative modeling of the dust around SN 2006bc proceeded using our 3D Monte Carlo RT code MOCASSIN (\citet{2005MNRAS.362.1038E}, and references therein).  Although we have presented evidence that 2006bc exploded asymmetrically, we do not have any strong observational constraints on the degree of asymmetry.  Consequently, the dust around SN 2006bc was modeled as a spherically, expanding shell with an inner radius, \textit{R$_{in}$}, and an outer radius, \textit{R$_{out}$}. We modeled the dust around SN 2006bc in two separate ways.  The first we refer to as our ``smooth" RT model.  Since the source of the luminosity is the ejecta, our smooth model assumes that the source luminosity is spread throughout the shell and proportional to the local density.  We further assumed a uniform dust distribution throughout the shell according to an \textit{r}$^{-2}$ density profile.   For our second model, we allow the dust to form into clumps.  For this reason, we call it our ``clumpy" model.  The clumpy model comprises an inhomogeneous distribution of spherical clumps embedded within an interclump medium of a supplied density.  The irradiating photons are only produced in this interclump medium with the clumps assumed to be dark.  The size of the spherical clumps are given by \textit{$r_{c}$ = 0.025 ($r_{out}$)} \citep{1994ApJ...425..814H,2006Sci...313..196S} with a volume filling factor, \textit{$f_{c}$} = 0.20, and density contrast \textit{$\alpha$ = $\rho_{c}$/$\rho$} = 1$\times$10$^{5}$.  For a more detailed discussion of modeling a clumpy environment with MOCASSIN see \citet{2007MNRAS.375..753E}.  For each model, we adopted a standard grain size distribution of $\textit{a}^{-3.5}$ between 0.005 and 0.05 $\mu$m \citep{2006Sci...313..196S, 2007ApJ...665..608M, 2009ApJ...704..306K}.  We tested two dust compositions with varying amounts of silicate (Si) and amorphous carbon (AC) dust.  The first was the AC dominated dust (75\% AC, 25\% Si), the second was Si dominated (25\% AC, 75\% Si).  Both models used the optical grain constants from \citet{1992A&A...261..567O} for Si and \citet{1988ioch.rept...22H} for AC.

For both the smooth and clumpy models, the primary input parameters that we varied were the ejecta temperature and luminosity, the inner and outer radii of the dust shell, and the mass of dust present.  Initial estimates of the ejecta luminosity and the extent of the dust shell, came from our optical photometry and the blackbody fits described above, respectively.  These parameters, along with the dust mass, were then varied to produce the best fit to the optical/$\textit{HST}$ and IRAC data points while simultaneously trying to predict more flux than the upper limits at 16 and 24 $\mu$m would allow.  

Our Si and AC dominated smooth models are both shown in Figure \ref{fig9}.  We found we were unable to predict sufficiently low flux at 24 $\mu$m for our silicate dominated models.  This is not unexpected as silicate dust has strong emission features at 10 and 18 $\mu$m which acts to hold up the SED flux relative to that at 3-8 $\mu$m.  Our best fit model, therefore, corresponds to dust with a composition of 75\% AC and 25\% silicate dust.  The other best fit parameters are an ejecta temperature of 5500 K, an ejecta luminosity of 1.2 x 10$^{6}$ L$_{\odot}$, a dust shell inner and outer radius of 67 AU and 6700 AU, respectively, and a total dust mass of 4 x 10$^{-4}$ M$_{\odot}$ (see Table 3).  This corresponds to a A$_{v}$ = 1.4.  If we allow the dust to form in clumps, the total amount of dust becomes $\sim$ 2.0 x 10$^{-3}$ M$_{\odot}$.  Even if all of this dust formed during the supernova explosion, it would still be orders of magnitude less than a typical core-collpase SN would need to form in order to explain the amount  of dust observed in the early universe \citep{2003A&A...409L..47B,2003MNRAS.343..427M}.

\subsection{Possible Light Echo}

The optical and near-IR lightcurve of SN 2006bc are presented in Figure \ref{fig10}.  After differenced imaging of our GMOS data, detections were made and magnitudes were measured on days 325, 359, and 413.  The data on day 532 and 694 are from \textit{HST}/WFPC2.  The photometry has not been corrected for foreground extinction.  SN 2006bc undergoes a steep drop immediately following maximum light leading us to categorize 2006bc as a Type II-L SN.  Assuming a total foreground extinction, A$_{\textit{v}}$ $\sim$ 1.6 (see above), and a distance to NGC 2397 of 20.5 Mpc, SN 2006bc reached a maximum brightness of \textit{V} $\sim$ -17.2.  Although showing normal peak brightness for a core-collapse SN, 2006bc then fades by $\sim$6 magnitudes during the first 400 days suggesting a relatively low $^{56}$Ni yield in the SN explosion \citep{2009ApJ...703.2205K} and/or an increase in extinction due to new dust.  Typically, such steep declines are seen in under-luminous SNe, though two Type II plateau SNe, 1994W \citep{1998ApJ...493..933S} and 2007od \citep{2010ApJ...715..541A}, were very highly luminous (\textit{V} $<$ -18) during the photospheric stage and yet showed low late-time luminosities attributed to a combination of low $^{56}$Ni mass and dust extinction.

By the onset of the nebular phase, SN 2006bc begins to decline with a rate in \textit{V} and \textit{I$_{c}$} that is consistent with the radioactive decay rate $^{56}$Co.  However, we observe an approximate 100 day flattening of the lightcurve in both \textit{V} and \textit{I$_{c}$} from days 416 to 541, after which a $^{56}$Co decline rate resumes.  The most likely explanation for this phenomenon is an optical light echo, and given the narrow time window in which the flattening was observed, we can place constraints on both the distance and geometry of the circumstellar material (CSM) responsible for the echo.  There are two obvious geometries that could explain the flattening of the lightcurve.  The first one has the CSM lying in a torus surrounding the SN.  In this case, due to light travel effects, the late onset and short duration of the flattening requires that the torus to be oriented close to face-on and have a thickness of no more than $\sim$ 100 lt-day.  Furthermore, the fact that flattening begins around day 413 suggests that the inner edge of the torus or ring must lie approximately 400 lt-day (1.04e18 cm) from the geometric center of the explosion.  This is consistent with the size of the main ring around 1987A.  

Another possible scenario is that the CSM is located within a thin spheroidal shell with a radius of approximately 400-500 lt-day.  Given the time it would take for light to travel to all parts of the shell and then be scattered in the direction to Earth, one is likely to assume that such a geometry could not produce the short duration flattening that we have observed.  However,  \citet{2005ApJS..159...60S} have shown the following proportion to be an empirically close fit to the average scattering function for MRN{\footnote{\citet{1977ApJ...217..425M}} dust  over scattering angles between $\theta=0$ to $\sim$ 160$\degr$.  The echo brightness scales as,   
\small
\begin{equation}
\textit{E.B.} \propto \left [ 1.0 +2.21\times10^{-3}\times(\theta)^{2} \right ]^{-1.123}.
\end{equation}
\normalsize 

This relation indicated that the echo brightness falls off sharply with increasing scattering angle, $\theta$.   With a thin shell of radius 400-500 lt-day, the scattered light from the CSM at low $\theta$ will begin to flatten out the lightcurve 400 to 500 days after the explosion while the material at increasing $\theta$ could continue to hold the lightcurve up for the subsequent 100 days.  Either of these scenarios are possible, though the latter seems more likely as the former requires such a specific inclination (\textit{i} $\gtrsim$ 80) of the torus to fit the data.  

\subsection{Conclusions}
We have presented multi-temporal, multi-wavelength observations of the core-collapse SN 2006bc during its nebular phase.  Based on an analysis of these observations we have drawn the following conclusions.

\begin{enumerate}

\item An analysis of the evolution of the optical spectra of SN 2006bc shows evidence that the average central wavelength of the SN H$\alpha$ emission profile is redshifted relative to the nebular emission from the galaxy between 325 to 663 days.  This is likely due to the asymmetric nature of the explosion.  The fact that this same redshift was not seen in the emission lines of [OI]  suggests the asymmetry was of the iron-peak elements only.  Furthermore, we find evidence for a gradual blue-shifting of the H$\alpha$ emission line profile during our observing period due to the formation of new dust within the SN ejecta.

\item Radiative transfer modeling of the dust around 2006bc has revealed an upper limit of 4$\times$10$^{-4}$ M$_{\odot}$ of new dust having formed during the SN explosion.  If we allow the dust to form into clumps, our models predict as much as 2 x 10$^{-3}$ M$_{\odot}$ of new dust formed during the explosion of SN 2006bc. These results are consistent with the mounting evidence suggesting dust formation within the ejecta of core-collapse SNe is insufficient to explain the large quantities of dust observed in the early universe.  

\item  An approximate 100 day flattening of the light curve beginning $\sim$ 400 days past explosion provides evidence for a light echo off a thin, spheroidal, shell of CSM surrounding SN 2006bc at a distance of ~ 400-500 lt-day.   

\end{enumerate}

\acknowledgments

This work has been supported by NSF grant AST-0707691 and HST grant HST-GO-11229.03-A. This work is based in part on observations made with the Spitzer Space Telescope, which is operated by the Jet Propulsion Laboratory, California Institute of Technology under a contract with NASA.  A portion of this data was obtained at the Gemini Observatory, which is operated by the Association of Universities for Research in Astronomy (AURA) under a cooperative agreement with the NSF on behalf of the Gemini partnership. The standard data acquisition has been supported by NSF grants AST-0503871 and AST-0803158 to A. U. Landolt.



{\it Facilities:} \facility{Gemini}, \facility{HST}, \facility{Spitzer}.

\bibliographystyle{apj}
\bibliography{gall0806}



\begin{figure}[!h]
\epsscale{1.0}
\plotone{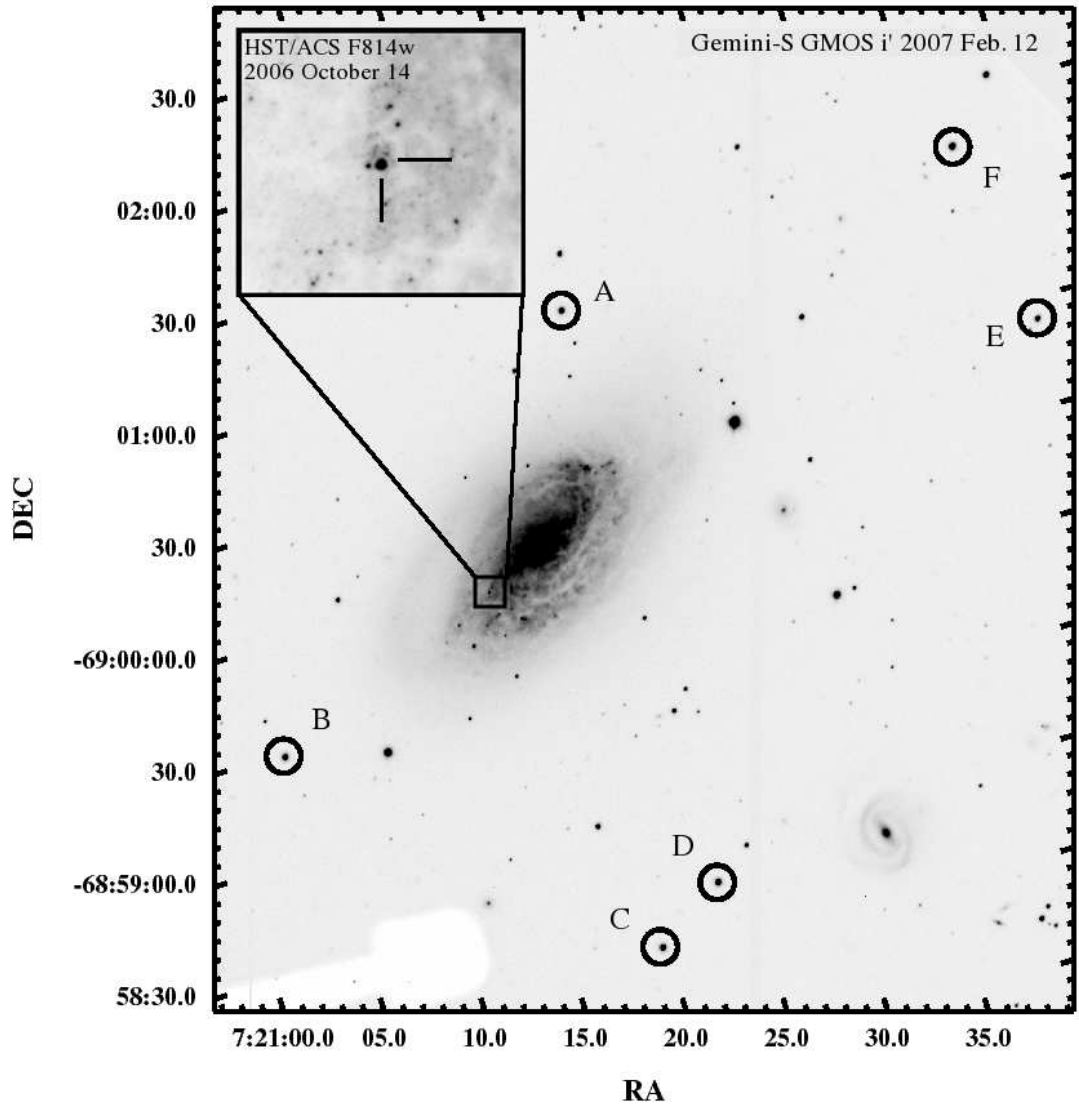}
\caption{Gemini/GMOS-S image of the field around NGC 2397 and SN 2006bc obtained 324 days after explosion with the \textit{i$^{\prime}$} filter.  A higher resolution inset showing the immediate region around SN 2006bc is shown and was obtained with \textit{HST/ACS} by \citet{2009MNRAS.395.1409S} as part of GO10498.  The figure also shows the secondary standard stars that we used to transform from the instrumental to standard magnitude system (See Table 4).\label{fig1}}
\end{figure}

\clearpage

\begin{figure}
\epsscale{1.0}
\plotone{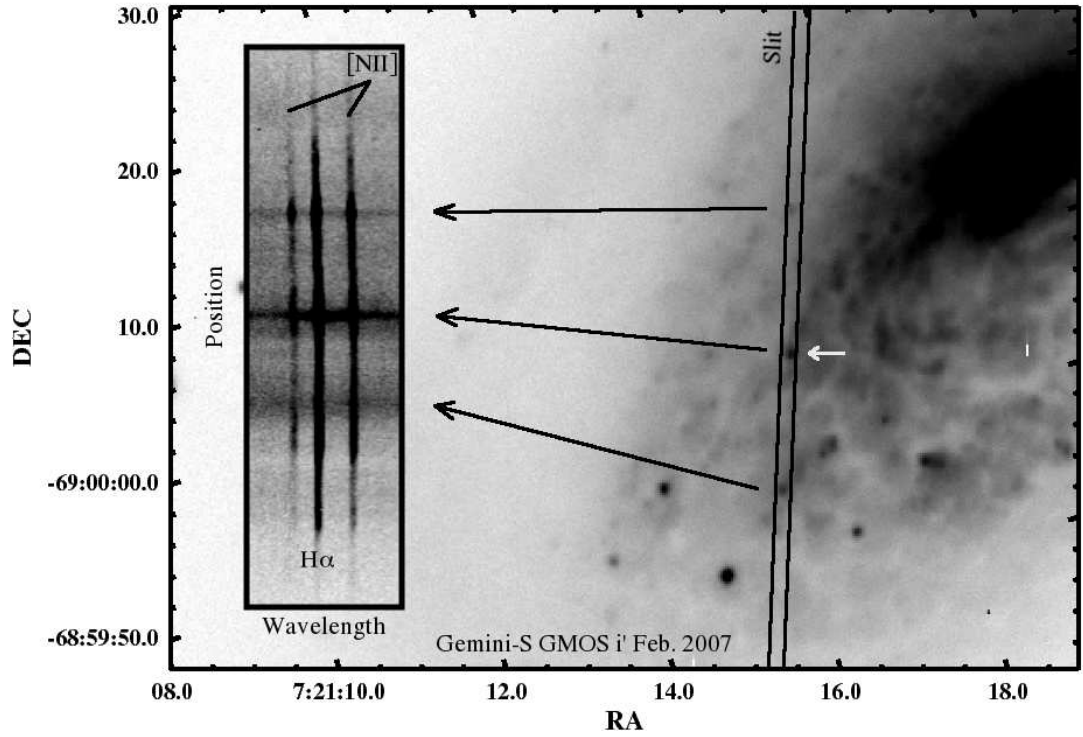}
\caption{GMOS-S \textit{i$^{\prime}$} image of SN 2006bc from 12 February 2007.  The position of the SN is marked with a white arrow.  The orientation of the slit for the February spectroscopic observations is depicted.  The image shows the crowded environment in which SN 2006bc is found, and the inset shows the section of the GMOS-S spectrum around H$\alpha$.  The respective origins of three emission regions along the spatial axis are marked with arrows in the image.  Care was taken to avoid emission from extraneous stars or HII regions (such as those shown above and below the SN position) in the section of sky used in the background subtraction.\label{fig2}}
\end{figure}


\begin{figure}
\begin{center}
\includegraphics[angle=0,scale=0.45]{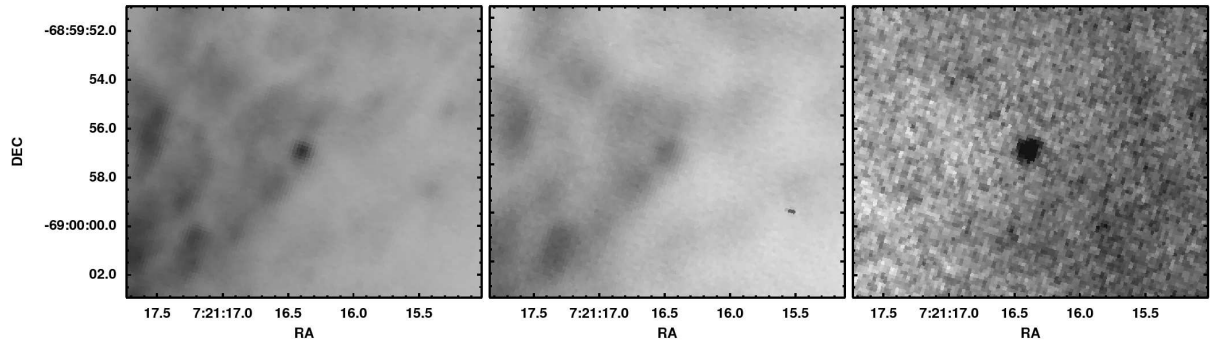}
\caption{Example PSF-matched differenced image from our GMOS data.  \textbf{Left}:  Day 324 i$^{\prime}$ image of SN 2006bc.  \textbf{Middle}: Day 662 i$^{\prime}$ image of SN 2006bc.  \textbf{Right}: Final subtracted image showing the SN with the nebular contamination considerably reduced.\label{fig3} }
\end{center}
\end{figure}

\begin{figure}
\begin{center}
\includegraphics[angle=90,scale=0.6]{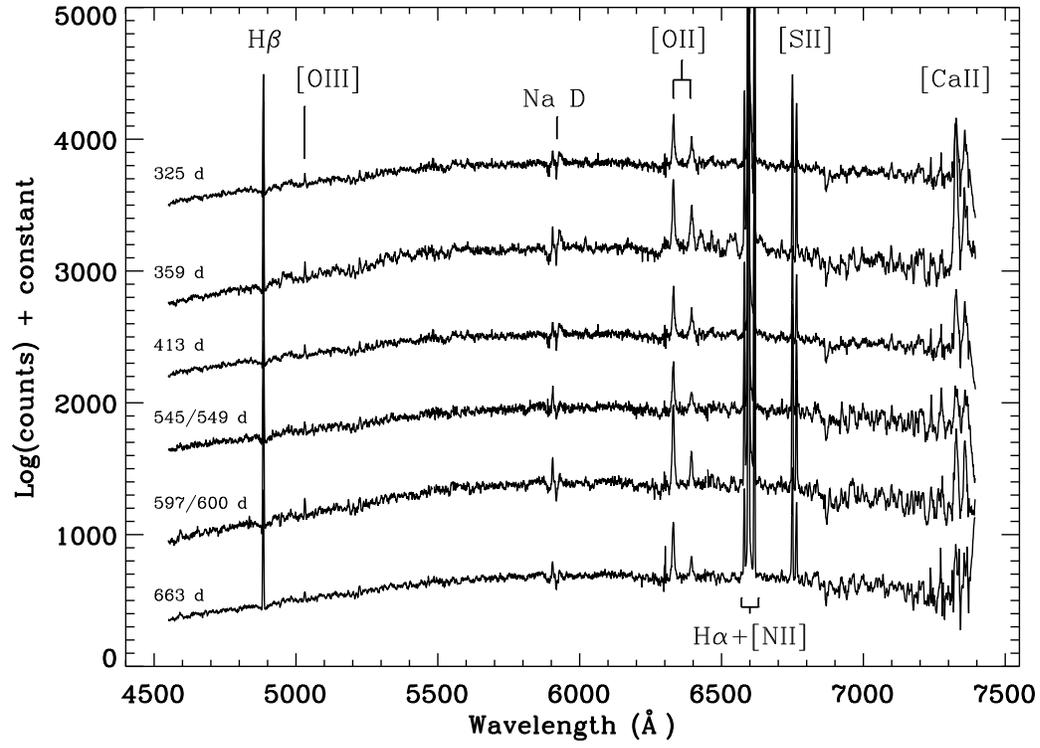}
\caption{Gemini/GMOS-S spectra SN 2006bc spanning 6 epochs.  Data have not been corrected for redshift or reddening. \label{fig4} }
\end{center}
\end{figure}

\begin{figure}
\begin{center}
\includegraphics[angle=90,scale=0.6]{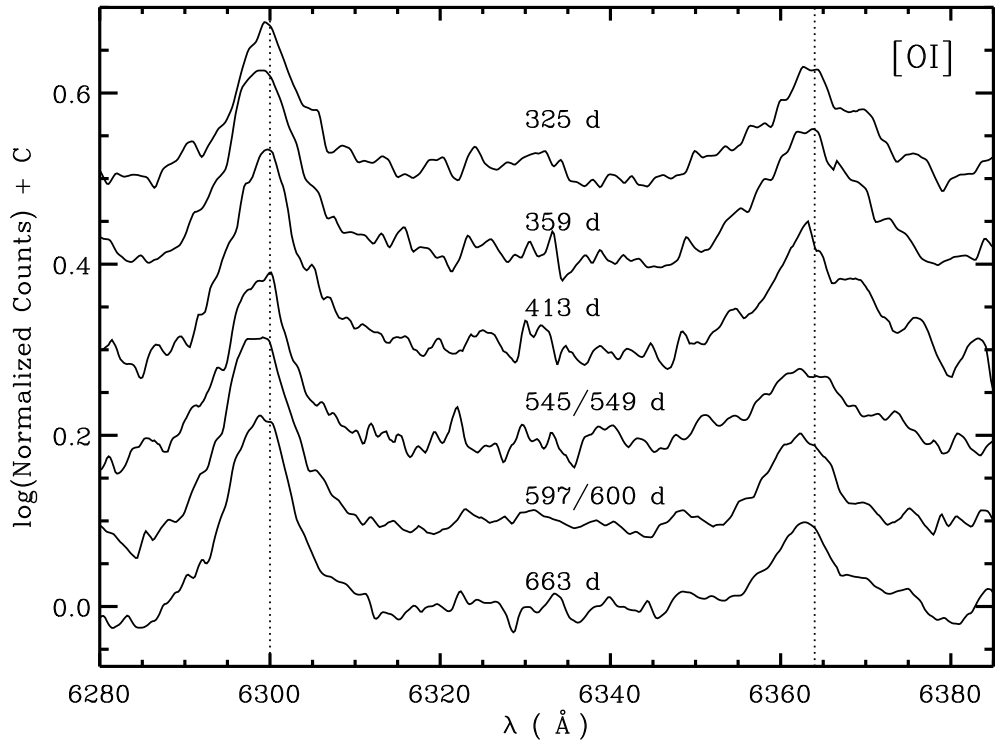}
\caption{Evolution of the [OI]$\lambda$6300,6363 emission lines across our six epochs.  The spectra have been redshift corrected, but they have not been dereddend. \label{fig5} }
\end{center}
\end{figure}

\begin{figure}
\begin{center}
\includegraphics[angle=90,scale=0.6]{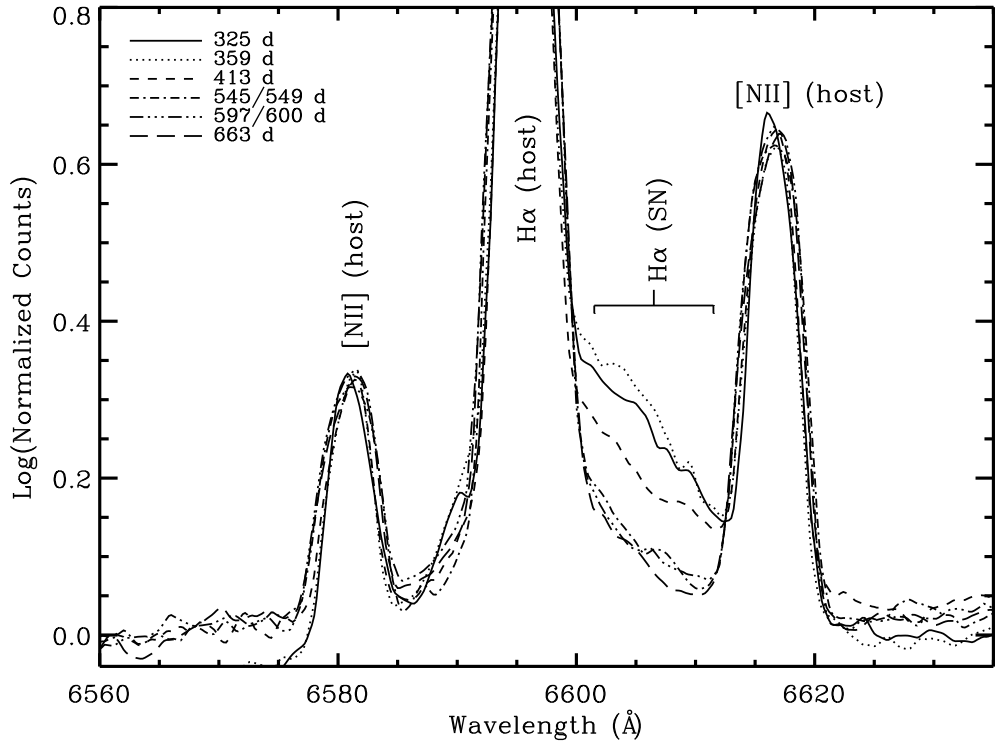}
\caption{Evolution of the H$\alpha$ + [NII] emission line profile of SN 2006bc.  Full profile shows strong contamination from narrow nebular emission.  This emission, originating in the host galaxy, is labeled as ``host."  Broad-width emission originating in the SN ejecta is also labeled. \label{fig6}}
\end{center}
\end{figure}

\begin{figure}
\epsscale{0.8}
\plotone{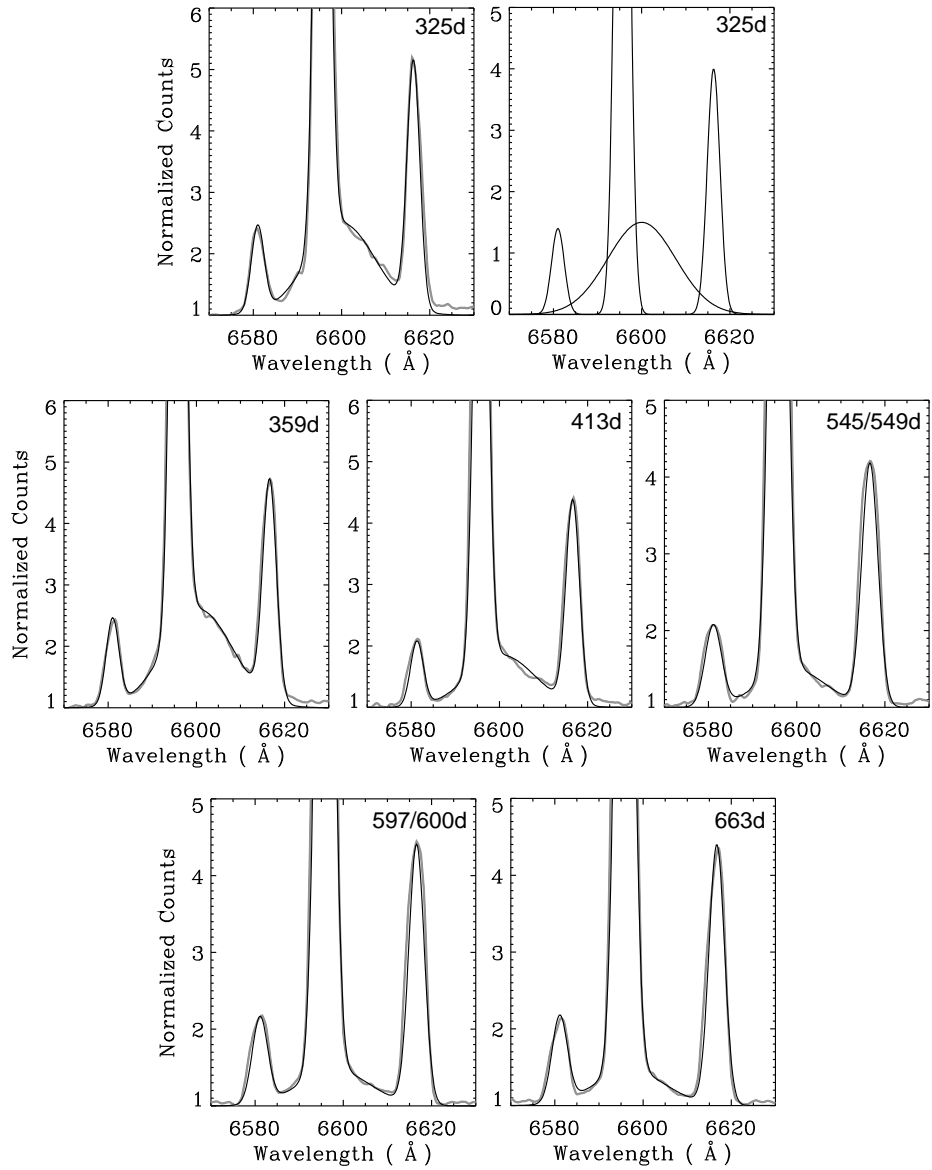}
\caption{Gaussian fits to the respective H$\alpha$ + [NII] profiles observed at each epoch.  The fits are shown in black while the data are shown in grey.  The fits are comprised of four Gaussian curves, one for each the SN ejecta  emission, the galactic [NII] $\lambda$6549 and $\lambda$6584 emission, and the galactic H$\alpha$ emission.  As an example, the four constituent parts of the fit for day 325 are given in the top-right panel.  The lower panels show the fits for subsequent epochs.\label{fig7}}
\end{figure}

\begin{figure}
\begin{center}
\includegraphics[angle=90,scale=0.60]{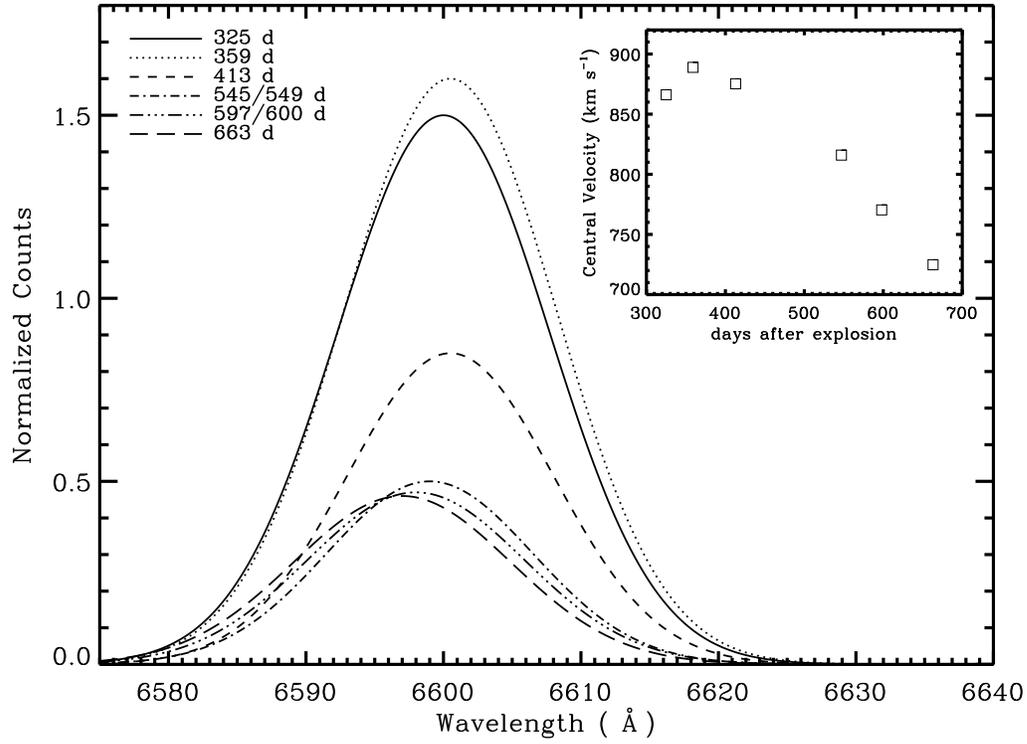}
\caption{Evolution of the SN component to our H$\alpha$ + [NII] fits from Figure \ref{fig7}.  Inset shows the central recessional velocity of the SN emission versus time.  The plot reveals an apparent blueshift of $\sim$200 km s$^{-1}$ of the SN emission profile from day 325 to 663. \label{fig8}}
\end{center}
\end{figure}

\begin{figure}
\begin{center}
\includegraphics[angle=0,scale=0.8]{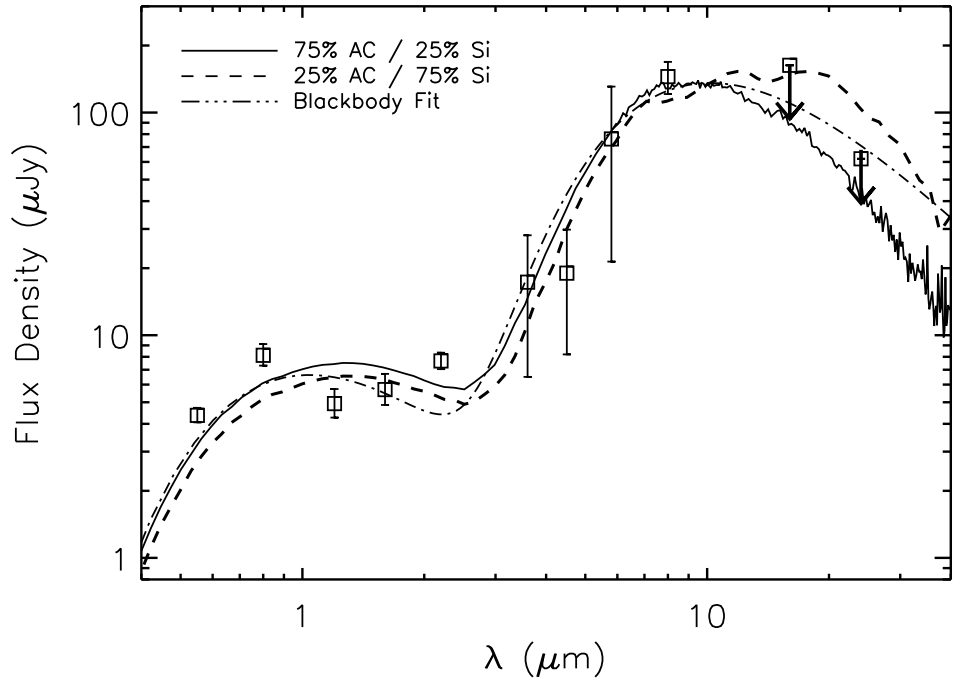}
\caption{Blackbody and MOCASSIN fits to the SED of SN 2006bc approximately 550 days past maximum light.  The optical data were taken on day 532 with \textit{HST}/WFPC2.  The JHK fluxes come from \textit{HST}/NICMOS data taken on day 540.  The 3.0, 4.5, 5.8, and 8.0 $\mu$m fluxes come from \textit{Spitzer}/IRAC data taken on day 537, while the 16 and 24 $\mu$m points were obtained with \textit{Spitzer} using IRS/PUI (day 496) and MIPS (day 688), respectively.  The dot-dash line is best blackbody fit to the data, the solid line is our best fit AC dominated, ``smooth" MOCASSIN radiative transfer model, and the dashed line is our best fit Si dominated,``smooth" model of the dust around SN 2006bc.  The data reveals a strong infrared excess by day 550.  Our AC dominated model represents the better fit to the data (particularly at 24 $\mu$), and predicts 4 x 10$^{-4}$ M$_{\odot}$ of dust around SN 2006bc.  If we allow for the dust to form into clumps, the total quantity of dust increases to 2 x 10$^{-3}$ M$_{\odot}$. \label{fig9}} 
\end{center}
\end{figure}

\begin{figure}
\begin{center}
\includegraphics[angle=-90,scale=0.60]{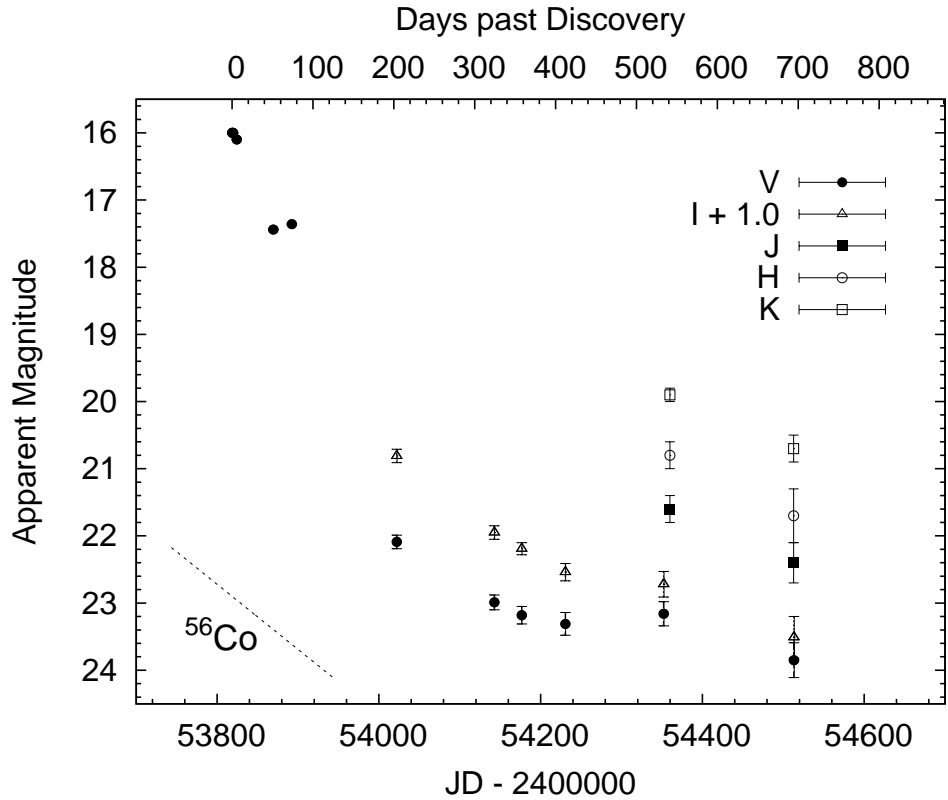}
\caption{\textit{VI$_{c}$JHK} lightcurve of SN 2006bc.   The data within the first 10 days are unfiltered discovery magnitudes while the data between day 40 and 60 comes from \textit{Astrosurf.com}.  The V and I data points from day 203 are measured from the ACS WFC images taken by \citet{2009MNRAS.395.1409S}.  The data beyond 300 days comes from the observations made for our study.  The lightcurves show an abrupt drop of $\sim$7-8 magnitude during the first 400 days past maximum light.  This is likely due to a combination of a  low $^{56}$Ni yield in the explosion and an increased optical extinction due to dust formation.  The data also show a short-lived flattening in the light curve from $\sim$400 to 500 days past maximum due to an optical light echo from CSM. \label{fig10}}
\end{center}
\end{figure}

\clearpage

\begin{deluxetable}{lccccc}
\tabletypesize{\small}
\tablecaption{Summary of Observations\label{tbl-1}}
\tablewidth{0pt}
\tablehead{
\colhead{Epoch$\tablenotemark{a}$} & \colhead{JD} & \colhead{Telescope} & \colhead{Instrument} & \colhead{Filters} & \colhead{Exp.Time (s)} 
}
\startdata
203$\tablenotemark{b}$ & 2454023 & \textit{HST} & ACS/WFC & F555W; F814W & 1500; 1600 \\
324 & 2454144 & Gemini South & GMOS-S spectra & -- & 3x900 \\
324 & 2454144 & Gemini South & GMOS-S imaging & \textit{g$^{\prime}$r$^{\prime}$i$^{\prime}$} & 60 \\ 
358 & 2454178 & Gemini South & GMOS-S spectra & -- & 3x900 \\
358 & 2454178 & Gemini South & GMOS-S imaging & \textit{g$^{\prime}$r$^{\prime}$i$^{\prime}$} & 60 \\
412 & 2454232 & Gemini South & GMOS-S spectra & -- & 3x900 \\
412 & 2454232 & Gemini South & GMOS-S imaging & \textit{g$^{\prime}$r$^{\prime}$i$^{\prime}$} & 60 \\
496 & 2454315 & \textit{Spitzer} & IRS/PUI  & Blue & 31\\
532 & 2454353 & \textit{HST} & WFPC2 & F606W; F814W & 1600; 1600 \\
537 & 2454356 & \textit{Spitzer} & IRAC & CH 1-4 & 10\\
540 & 2454359 & \textit{Spitzer} & MIPS & 24 $\mu$m & 31 \\
541 & 2454361 & \textit{HST} & NICMOS & F110W; F160W; F205W & 640; 512; 576\\
544 & 2454364 & Gemini South & GMOS-S spectra & -- & 3x1050 \\
544 & 2454364 & Gemini South & GMOS-S imaging & \textit{g$^{\prime}$r$^{\prime}$i$^{\prime}$} & 60 \\
596 & 2454416 & Gemini South & GMOS-S spectra & -- & 3x1050 \\
596 & 2454416 & Gemini South & GMOS-S imaging & \textit{g$^{\prime}$r$^{\prime}$i$^{\prime}$} & 60 \\
662 & 2454482 & Gemini South & GMOS-S spectra & -- & 3x1050 \\
662 & 2454482 & Gemini South & GMOS-S imaging & \textit{g$^{\prime}$r$^{\prime}$i$^{\prime}$} & 60 \\
648 & 2454467 & \textit{Spitzer} & IRS/PUI & Blue & 31\\
688 & 2454507 & \textit{Spitzer} & MIPS & 24 $\mu$m & 31 \\
690 & 2454509 & \textit{Spitzer} & IRAC & CH 1-4 & 10 \\
694 & 2454514 & \textit{HST} & WFPC2 & F606W; F814W & 1600; 1600\\
753 & 2454572 & \textit{Spitzer} & MIPS & -- & 31 \\
776 & 2454595 & \textit{Spitzer} & IRAC & CH 1-4 & 27
\enddata
\tablenotetext{a}{Assuming explosion date JD $=$ 2453819}
\tablenotetext{b}{Data taken as part of GO10498 \citep{2009MNRAS.395.1409S}.  Data were acquired from the archive, re-reduced and analyzed.}
\end{deluxetable}

\begin{deluxetable}{ll}
\tabletypesize{\small}
\tablecaption{Near and Mid-IR fluxes of SN 2006bc for day $\sim550$\label{tbl-2}}
\tablewidth{0pt}
\tablehead{
\colhead{$\lambda$ ($\mu$m)} & \colhead{Flux ($\mu$Jy)}
}
\startdata
1.2   &    4.9$^{+0.7}_{-0.8}$\\
1.6   &  5.7$^{+0.8}_{-1.0}$  \\
2.2   &  7.7$^{+0.6}_{-0.7}$\\  
3.6   &  17.3$^{+10.8}_{-10.8}$  \\
4.5   &  19.0$^{+10.8}_{-10.8}$ \\
5.8   &  76.1$^{+57.5}_{-57.5}$\\  
8.0    & 145.0$^{+23.9}_{-23.9}$ \\
16.0  & $<$ 163.0\\
24.0   & $<$ 62.0
\enddata
\end{deluxetable}

\clearpage


\begin{deluxetable}{lcccccccc}
\tabletypesize{\small}
\tablecaption {Best Fit Monte Carlo Radiative Transfer Models\label{tbl-4}} 
\tablewidth{0pt}
\tablehead{
\colhead{Model} & 
\colhead{AC/Si} & 
\colhead{T (K)} & 
\colhead{R$_{in}$ (AU)} & 
\colhead{R$_{out}$(AU)} & 
\colhead{L (L$_{\odot}$)} & 
\colhead{$\tau$$_{v}$} & 
\colhead{M$_{d}$ (M$_{\odot}$)}
}
\startdata
Smooth & 0.75/0.25 & 5500 & 67 & 6700 & 1.2e6 & 1.27 & 4e-4\\
Clumpy & 0.75/0.25 & 5500 & 67 & 2000 & 1.2e6 & 1.00 & 2e-3
\enddata
\end{deluxetable}

\begin{deluxetable}{lccc}
\tabletypesize{\small}
\tablecaption{Secondary Standard Stars\label{tbl-5}}
\tablewidth{0pt}
\tablehead{
\colhead{ID} & \colhead{V ($\delta$V)} & \colhead{R ($\delta$R)} & \colhead{I ($\delta$I)}
}
\startdata
A & 17.140 (0.009) & 16.541 (0.006) & 16.032 (0.023) \\
B & 17.003 (0.008) & 16.367 (0.010) & 15.826 (0.011) \\
C & 16.548 (0.013) & 16.118 (0.009) & 15.723 (0.051) \\
D & 16.878 (0.012) & 16.180 (0.013) & 15.624 (0.017) \\
E & 17.618 (0.011) & 17.050 (0.009) & 16.539 (0.106) \\
F & 16.348 (0.011) & 16.014 (0.005) & 15.656 (0.010) 
\enddata
\end{deluxetable}

\end{document}